\documentclass[prd,aps,nofootinbib,showpacs,10pt]{revtex4}
\usepackage{amsmath,graphicx,color,epsfig}

\def\pp{{\prime\prime}}
\def\vp{\varepsilon}

\begin{document}
\title{$B$ to tensor meson form factors  in the perturbative QCD approach}
\author{  Wei Wang }

\affiliation{
 \it  Istituto Nazionale di Fisica Nucleare, Sezione di Bari, Bari 70126, Italy}

\begin{abstract}
We calculate the $B_{u,d,s}\to T$ form factors within the framework
of the perturbative QCD approach, where $T$ denotes a light tensor
meson with $J^P=2^+$. Due to the similarities between the wave
functions of a vector and a tensor meson, the factorization formulas
of $B\to T$ form factors can be obtained from the $B\to V$
transition through a replacement rule. As a consequence, we find
that these two sets of form factors have the same signs and
correlated $q^2$-dependence behaviors. At $q^2=0$ point, the $B\to
T$ form factors are smaller than the $B\to V$ ones, in accordance
with the experimental data of radiative $B$ decays.  In addition, we
use our results for the form factors to explore semilteptonic $B\to
Tl\bar \nu_l$ decays and the branching fractions can reach the order
$10^{-4}$.
\end{abstract}
 \maketitle
\section{Introduction}\label{section:introduction}

In the quark model, a meson is composed of one quark pair and the
spin-parity quantum numbers $J^{P}$ of a meson state is consequently
fixed by this constituent quark pair, for instance $J^P=0^+$ for
pseudoscalar mesons. For p-wave tensor mesons with $J^P=2^+$, both
orbital angular momentum $L$ and the total spin $S$ of the quark
pair are equal to 1. By making use of the flavor SU(3) symmetry, the
nine mesons, isovector mesons $a_2(1320)$, isodoulet states
$K_2^*(1430)$ and two isosinglet mesons $f_2(1270),f_2'(1525)$, form
the first $^3P_2$ nonet~\cite{Amsler:2008zz}. These mesons have been
well established in various processes.

$B$ meson decays into tensor mesons are of prime interest in several
aspects. The main experimental observables in hadronic $B$ decays,
branching ratios and CP asymmetries, are helpful to inspect
different theoretical computations. One exploration concerns the
isospin symmetry. For instance the $B\to K_2^*(1430)\eta$ channel
has already been observed in 2006 with the branching ratio (BR)
$(9.1\pm3.0)\times 10^{-6}$ for the charged channel and a similar
one $(9.6\pm2.1)\times 10^{-6}$ for the neutral
channel~\cite{Aubert:2006fj}. But the $B\to\omega K_2^*(1430)$ mode
possesses a large isospin violation: the BR for the neutral mode
$(10.1\pm2.3)\times 10^{-6}$ is about one half of that for the
charged mode $ (21.5\pm4.3)\times 10^{-6}$~\cite{Aubert:2009sx}.
Moreover, polarizations of the final mesons in $B$ decays can shed
light on the helicity structure of the electroweak interactions. In
the standard model, they are expected to obey a specific hierarchy
when factorization is adopted to handle the decay amplitudes and the
heavy quark symmetry is exploited to derive relations among the
involved form factors. In particular the longitudinal polarization
fraction is expected to be close to unity. Deviations from this rule
have already been experimentally detected in several $B$ decays to
two light vector mesons, with the implication of something beyond
the naive expectation. Towards this direction $B$ meson decays into
a tensor meson can play a complementary role. For example the decay
mode $B\to\phi K_2^*(1430)$ is mainly dominated by the longitudinal
polarization~\cite{:2008zzd,Aubert:2008zza}, in contrast with the
$B\to \phi K^*$ where the transverse polarization is comparable with
the longitudinal one~\cite{HFAG}.

Despite a number of interesting decay modes have been detected on
the experimental side, currently there exist few theoretical
investigations on $B$ to tensor transitions. Since a tensor meson
can not be produced by a local vector or axial-vector current, the
$B\to MT$ decay amplitude is reduced  in terms of the $B\to T$
transition and the emission of a light meson $M$. The motif of this
work is to handle the first sector with the computation of the $B\to
T$ form factors, and in particular we will use the perturbative QCD
(PQCD) approach \cite{Keum:2000ph} which is based on the $k_T$
factorization. If the recoiling meson in the final state moves very
fast, a hard gluon is required to kick the soft light quark in $B$
meson into an energetic one and then the process is perturbatively
calculable. Keeping quarks' intrinsic transverse momentum, the PQCD
approach is free of endpoint divergence and the Sudakov formalism
makes it more self-consistent. As a direct consequence, we can do
the form factor calculation and the quantitative annihilation type
diagram calculation in this approach. Our results for these form
factors in this work will serve as necessary inputs in the future
analysis of the semileptonic and nonleptonic $B$ decays into a
tensor meson.

This paper is organized as follows. In Sec.~II, we collect the input
quantities, including the $B$-meson wave function, light-cone
distribution amplitudes (LCDAs) of light tensor mesons. In
Sec.~\ref{section:formfactor}, we discuss the factorization property
of the $B\to T$ form factors in the PQCD approach. Subsequently we
present our numerical results and a comparison with other   model
predictions is also given. Branching ratios, polarizations and
angular asymmetries of the semileptonic $B\to Tl\bar\nu_l$ decays
are predicted in Sec.~\ref{sec:semileptonic}. Our summary is given
in the last section.
\section{Wave functions}\label{section:LCDAs}
%

We will work in the $B$ meson rest frame  and employ the light-cone
coordinates for momentum variables. In the heavy quark limit the
light tensor meson in the final state moves very fast in the
large-recoil region, we choose its momentum mainly on the plus
direction in the light-cone coordinates. The momentum of $B$ meson
and the light meson can be written as
 \begin{eqnarray}
 P_{B}=\frac{m_{B}}{\sqrt{2}}(1,1,0_{\perp})\;,\;
 P_2=\frac{m_{B}}{\sqrt{2}}(\eta,\frac{r_2^2}{\eta},0_{\perp})\;,\label{eq:momentum}
 \end{eqnarray}
where $r_2\equiv \frac{m_{T}}{m_{B}}$, with $m_{T},m_B$ as the mass
of the tensor meson and the $B$ meson, respectively.  The
approximate relation $\eta\approx1-q^2/m_{B}^2$ holds for the
momentum transfer $q=P_{B}-P_2$. The momentum of the light antiquark
in $B$ meson and the quark in light mesons are denoted as $k_1$ and
$k_2$ respectively
 \begin{eqnarray}
 k_1=(0,\frac{m_{B}}{\sqrt{2}}x_1,\textbf{k}_{1\perp})\;,\;k_2=(\frac{m_{B}}{\sqrt{2}}
 x_2\eta,0,\textbf{k}_{2\perp})\;,\label{eq:fmomentum}
 \end{eqnarray}
with $x_i$ being the momentum fraction.

The spin-2 polarization tensor, which satisfies $\epsilon_{\mu\nu}
P^{\nu}_2=0$, is symmetric and traceless. It  can be constructed via
the spin-1 polarization vector $\epsilon$:
\begin{eqnarray}
 &&\epsilon_{\mu\nu}(\pm2)=
 \epsilon_\mu(\pm)\epsilon_\nu(\pm),\;\;\;\;
 \epsilon_{\mu\nu}(\pm1)=\frac{1}{\sqrt2}
 [\epsilon_{\mu}(\pm)\epsilon_\nu(0)+\epsilon_{\nu}(\pm)\epsilon_\mu(0)],\nonumber\\
 &&\epsilon_{\mu\nu}(0)=\frac{1}{\sqrt6}
 [\epsilon_{\mu}(+)\epsilon_\nu(-)+\epsilon_{\nu}(+)\epsilon_\mu(-)]
 +\sqrt{\frac{2}{3}}\epsilon_{\mu}(0)\epsilon_\nu(0).
\end{eqnarray}
In the case of the tensor meson moving on the plus direction of the
$z$ axis, the explicit structures of $\epsilon$ in the ordinary
coordinate frame are chosen as
\begin{eqnarray}
\epsilon_\mu(0)&=&\frac{1}{m_T}(|\vec p_T|,0,0,E_T),\;\;\;
\epsilon_\mu(\pm)=\frac{1}{\sqrt{2}}(0,\mp1,-i,0),
\end{eqnarray}
where $E_T$ and $\vec{p}_T$ is the energy and the magnitude of the
tensor meson momentum in the $B$ rest frame, respectively. In the
following calculation, it is convenient to introduce a new
polarization vector $\epsilon_T$ for the involved tensor meson
\begin{eqnarray}
  &&\epsilon_{T\mu}(h) =\frac{1}{m_B}
  \epsilon_{\mu\nu}(h)P_{B}^\nu,
\end{eqnarray}
which satisfies
\begin{eqnarray}
  && \epsilon_{T\mu}(\pm2)=0,\;\;\;
  \epsilon_{T\mu}(\pm1)=\frac{1}{m_B}\frac{1}{\sqrt2}\epsilon(0)\cdot
  P_{B}\epsilon_\mu(\pm),\;\;\;
  \epsilon_{T\mu}(0)=\frac{1}{m_B}\sqrt{\frac{2}{3}}\epsilon(0)\cdot
  P_{B}\epsilon_\mu(0).
\end{eqnarray}
The contraction is evaluated as $\epsilon(0)\cdot P_{B}/m_B=|\vec
p_T|/m_T$ and thus we can see that the new vector $\epsilon_T$ plays
a similar role with the ordinary polarization vector $\epsilon$,
regardless of the dimensionless constants $\frac{1}{\sqrt2}|\vec
p_T|/m_T$ or $\sqrt{\frac{2}{3}}|\vec p_T|/m_T$.

Tensor meson decay constants are defined through matrix elements of
local current operators between the vacuum and a meson
state~\cite{Cheng:2010hn}
\begin{equation}
\langle T|j_{\mu\nu}(0)|0\rangle=f_Tm_T^2\epsilon_{\mu\nu}^*,\;\;\;
\langle T|j_{\mu\nu\rho}|0\rangle =-if^T_T
m_T(\epsilon_{\mu\delta}^* P_{2\nu}-\epsilon_{\nu\delta}^*
P_{2\mu}).
\end{equation}
The interpolating current for $f_T$ is chosen as
$j_{\mu\nu}=\frac{1}{2} [\bar q_1(0)\gamma_\mu i
\buildrel\leftrightarrow\over D_\nu q_2(0)+\bar q_1(0)\gamma_\nu i
\buildrel\leftrightarrow\over D_\mu q_2(0)]$ with the covariant
derivative $ \buildrel\leftrightarrow\over D_\nu =
\buildrel\rightarrow\over D_\nu - \buildrel\leftarrow\over D_\nu $:
$\buildrel\rightarrow\over D_\nu=\buildrel\rightarrow\over
\partial_\nu+ig_sA^a_\nu\lambda^a/2$ and
$\buildrel\leftarrow\over D_\nu=\buildrel\leftarrow\over
\partial_\nu-ig_sA^a_\nu\lambda^a/2$; the one for $f_T^\perp$ is selected as $
 j_{\mu\nu\rho}^\dagger= \bar q_2(0)\sigma_{\mu\nu} i\buildrel\leftrightarrow\over
 D_\delta(0) q_1(0).$
These quantities have been partly calculated in the QCD sum rules in
Refs.~\cite{Aliev:1981ju,Aliev:1982ab,Aliev:2009nn} and we quote the
recently updated results from Ref.~\cite{Cheng:2010hn} in
Tab.~\ref{Table:Tdecayconstant}. One interesting feature in these
values is that the two decay constants of $a_2(1320)$ are almost
equal but large differences are found for $K_2^*(1430)$ and
$f_2'(1525)$. In the case of a light vector meson, taking $\rho$ as
an example, the transverse decay constant  is  typically about
(20\%--30\%) smaller than the longitudinal one:
$f_\rho^T/f_\rho=(0.687\pm0.027)$~\cite{Allton:2008pn}.

\begin{table}
\caption{Decay constants  of tensor mesons from
Ref.~\cite{Cheng:2010hn} (in units of MeV)}
\begin{tabular}{cccccccccc}
\hline\hline
   $f_{a_2} $   & $ f_{a_2}^T $
 & $ f_{K_2^*} $ & $ f_{K_2^*}^T $  & $f_{f_2(1270)} $ & $f_{f_2(1270)}^T$
   & $f_{f_2'(1525)} $ & $f_{f_2'(1525)}^T$ \\
\ \ \
   $107\pm6$ & $105\pm 21$    & $ 118\pm 5$  & $77\pm 14$
\ \ \
 & $102\pm 6$ & $117\pm25$    & $126\pm 4$  & $65\pm 12$\\
\hline \hline
\end{tabular}\label{Table:Tdecayconstant}
 \end{table}

In the PQCD approach, the necessary inputs contain the LCDAs which
are constructed by matrix elements of the non-local operators at the
light-like separations $z_\mu$ with $z^2=0$, and sandwiched between
the vacuum and the meson state. For tensor mesons, their
distribution amplitudes are recently analyzed in
Ref.~\cite{Cheng:2010hn} which will provide a solid foundation in
our study of $B\to T$ form factors. The LCDAs up to twist-3 for a
generic tensor meson are defined by:
\begin{eqnarray}
\langle T(P_2,\epsilon)|\bar q_{2\beta}(z) q_{1\alpha} (0)|0\rangle
&=&\frac{1}{\sqrt{2N_c}}\int_0^1 dx e^{ixP_2\cdot z} \left[m_T\not\!
\epsilon^*_{\bullet L} \phi_T(x) +\not\! \epsilon^*_{\bullet
L}\not\! P_2 \phi_{T}^{t}(x) +m_T^2\frac{\epsilon_{\bullet} \cdot
n}{P_2\cdot n} \phi_T^s(x)\right]_{\alpha\beta},
\nonumber\\
 \langle T(P_2,\epsilon)|\bar q_{2\beta}(z) q_{1\alpha}
(0)|0\rangle &=&\frac{1}{\sqrt{2N_c}}\int_0^1 dx e^{ixP_2\cdot z}
\left[ m_T\not\! \epsilon^*_{\bullet T}\phi_T^v(x)+
\not\!\epsilon^*_{\bullet T}\not\! P_2\phi_T^T(x)+m_T
i\epsilon_{\mu\nu\rho\sigma}\gamma_5\gamma^\mu\epsilon_{\bullet
T}^{*\nu} n^\rho v^\sigma \phi_T^a(x)\right ]_{\alpha\beta}\;,
\label{spf}
\end{eqnarray}
for the longitudinal polarization ($h=0$) and transverse
polarizations ($h=\pm1$), respectively.  Here $x$ is the momentum
fraction associated with the $q_2$ quark. $n$ is the moving
direction of the vector meson and $v$ is the opposite direction.
$N_c=3$ is the color factor and  the convention $\epsilon^{0123}=1$
has been adopted. The new vector $\epsilon_\bullet$ in
Eq.~\eqref{spf} is related to the polarization tensor by
$\epsilon_{\bullet\mu}\equiv\frac{\epsilon_{\mu\nu} v^\nu}{P_2\cdot
v}m_T=\frac{2\epsilon_{\mu\nu} P_B^\nu}{m_B^2-q^2}m_T$ and moreover
it plays the same role with the polarization vector $\epsilon$  in
the definition of the vector meson LCDAs. The above distribution
amplitudes can be related to the ones given in
Ref.~\cite{Cheng:2010hn} by~\footnote{The distribution amplitudes
$h_{||}^{(s)},h_{||}^{(t)}$ and $g_{\perp}^{(v)}$ correspond to
$h_s,h_t$ and $g_v$ in Ref.~\cite{Cheng:2010hn}, but our
$g^{(a)}_\perp$ differs from their $g_a$ by a factor of 2:
$g_\perp^{(a)}=2g_a$. This definition is more convenient in the
following analysis of form factors. }
\begin{eqnarray}
&&\phi_{T}(x)=\frac{f_{T}}{2\sqrt{2N_c}}\phi_{||}(x),\;\;\;
\phi_{T}^t(x)=\frac{f_{T}^T}{2\sqrt{2N_c}}h_{||}^{(t)}(x),\nonumber\\
&&\phi_{T}^s(x)=\frac{f_{T}^T}{4\sqrt{2N_c}}
\frac{d}{dx}h_{||}^{(s)}(x),\hspace{3mm}
\phi_{T}^T(x)=\frac{f_{T}^T}{2\sqrt{2N_c}}\phi_{\perp}(x)
,\nonumber\\
&&\phi_{T}^v(x)=\frac{f_{T}}{2\sqrt{2N_c}}g_{\perp}^{(v)}(x),
\hspace{3mm}\phi_{T}^a(x)=\frac{f_{T}}{8\sqrt{2N_c}}
\frac{d}{dx}g_{\perp}^{(a)}(x).
\end{eqnarray}

The twist-2 LCDA can be expanded in terms of Gegenbauer polynomials
$C_n^{3/2}(2x-1)$ weighted by the  Gegenbauer moments. Particularly
its asymptotic form is
\begin{eqnarray}
 \phi_{||,\perp} (x)&=& 30x(1-x)(2x-1),\label{phiV}
\end{eqnarray}
with the normalization conditions
\begin{eqnarray}
 \int_0^1 dx (2x-1) \phi_{||,\perp} (x)=1.
\end{eqnarray}

Using equation of motion in QCD, two-particle twist-3 distribution
amplitudes are expressed as functions of the twist-2 LCDAs and the
three-particle twist-3 LCDAs. In the Wandzura-Wilczek limit, i.e.
with the neglect of the three-particle terms,  the asymptotic forms
of twist-3 LCDAs are derived as~\cite{Cheng:2010hn}
\begin{eqnarray}
h_\parallel^{(t)}(x) & = & \frac{15}{2}(2x-1)(1-6x+6x^2) ,\;\;\;
h_{||}^{(s)}(x)   = 15x(1-x)(2x-1),\\
g_\perp^{(a)}(x) & = & 20x(1-x)(2x-1) ,\;\;\; g_\perp^{(v)}(x)
=5(2x-1)^3.
\end{eqnarray}

%
%
%
%
%

Since the $B$ meson is a pseudoscalar heavy meson, the two structure
$(\gamma^\mu\gamma_5)$ and $\gamma_{5}$ components remain as leading
contributions. Then, $\Phi_{B}$ is written by
\begin{equation}
 \Phi_{B} = \frac{i}{\sqrt{6}}
\left\{ (\not \! P_B \gamma_5) \phi_B^A + \gamma_{5} \phi_B^P
\right\},
\end{equation}
where $\phi_B^{A,P}$ are Lorentz scalar distribution amplitudes. As
shown in Ref.~\cite{Lu:2002ny}, $B$ meson's wave function can be
simplified into
\begin{equation}
 \Phi_{B}(x,b) = \frac{i}{\sqrt{6}}
\left[ (\not \! P_B \gamma_5) + m_B \gamma_5 \right] \phi_B(x,b),
\end{equation}
where the numerically-suppressed terms in the PQCD approach have
been neglected. For the distribution amplitude, we adopt the model:
\begin{eqnarray}
\phi_{B}(x,b)=N_{B}x^{2}(1-x)^{2}\exp \left[ -\frac{1}{2} \left(
\frac{xm_{B}}{\omega _{B}}\right) ^{2} -\frac{\omega
_{B}^{2}b^{2}}{2}\right] \label{bw} \;,
\end{eqnarray}
with  $\omega_{B}$ being the shape parameter and $N_B$ as the
normalization constant. In the above parametrization form, $\phi_B$
will have a sharp peak at $x\sim 0.1$, in accordance with the most
probable momentum fraction of the light quark: $\Lambda_{\rm
QCD}/m_B$. Here $\Lambda_{\rm QCD}$ denotes the typical
hadronization scale. In recent years, a number of studies of
$B^{\pm}$ and $B_d^0$ decays have been performed in the PQCD
approach, from which the $\omega_b$ is found around
$0.40\mbox{GeV}$~\cite{Keum:2000ph,Lu:2002ny}. In our calculation,
we will adopt $\omega_b=(0.40\pm0.05)\mbox{GeV}$ and
$f_B=(0.19\pm0.02)\rm{GeV}$ for $B$ mesons. For the $B_s$ meson,
taking the SU(3) breaking effects into consideration, we employ
$\omega_{b}=(0.50\pm 0.05)\mbox{GeV}$~\cite{Ali:2007ff} and
$f_{B_s}=(0.23\pm0.02)\rm{GeV}$. These values for decay constants
are consistent with the recent Lattice QCD
simulations~\cite{Gamiz:2009ku}
\begin{eqnarray}
 f_{B}=(0.190\pm0.01) {\rm GeV},\;\;\;
 f_{B_s}=(0.231\pm0.015) {\rm GeV}.
\end{eqnarray}

\section{ $B\to T$  form factors in the PQCD approach}\label{section:formfactor}

\subsection{PQCD approach}

\begin{figure}[tb]
\begin{center}
\psfig{file=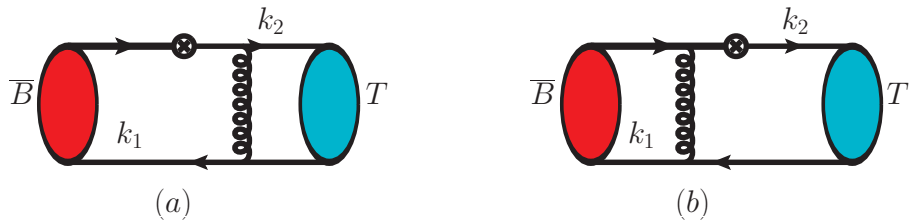,width=12.0cm,angle=0}
\end{center}
\caption{Feynman diagrams of $B$ meson decays into a tensor meson.
The cross represents the weak current, from which a lepton pair can
be emitted.}\label{fig:transition}
\end{figure}

The most important feature of the PQCD approach is that it takes
into account the intrinsic transverse momentum of valence quarks.
The tree-level transition amplitude, taking the first diagram in
Fig.~\ref{fig:transition} as an example, can be directly expressed
as a convolution of wave functions $\phi_B$, $\phi_2$ and hard
scattering kernel $T_H$ with both longitudinal momenta and
transverse space coordinates
\begin{eqnarray}
{\cal M}=\int^1_0dx_1dx_2\int
{d^2{\vec{b}}_{1}}{d^2{\vec{b}}_{2}}{\phi}_B(x_1,{\vec{b}}_{1},P_B,t)
T_H(x_1,x_2,{\vec{b}}_{1},{\vec{b}}_{2},t){\phi}_2(x_2,{\vec{b}}_{2},P_2,t).
\end{eqnarray}
Individual higher order diagrams may suffer from two generic types
of infrared divergences: soft and collinear. In both cases, the loop
integration generates logarithmic divergences. These divergences can
be separated from the hard kernel and reabsorbed into meson wave
functions using eikonal approximation~\cite{Li:1994iu}. When soft
and collinear momentum overlap, double logarithm divergences will be
generated and they can be grouped into the Sudakov factor using the
technique of resummation.  In the threshold region, loop corrections
to the weak decay vertex will also produce double logarithms which
can be factored out from the hard part and grouped into the quark
jet function. Resummation of the double logarithms results in the
threshold factor $S_t$~\cite{Li:2001ay}. This factor decreases
faster than any other power of ${x}$ as ${x\rightarrow 0}$, which
modifies the behavior in the endpoint region to make PQCD approach
more self-consistent. For a review of this approach, please see
Ref.~\cite{Li:2003yj}.

With the inclusion of the Sudakov factors, we can get the generic
factorization formula in the PQCD approach:
\begin{eqnarray}
{\cal M}&=&\int^1_0dx_1dx_2\int
{d^2{\vec{b}}_{1}}{d^2{\vec{b}}_{2}} {\phi}_B(x_1,{\vec{b}}_{1},P_B,t)
T_H(x_1,x_2,{\vec{b}}_{1},{\vec{b}}_{2},t){\phi}_2(x_2,{\vec{b}}_{2},P_2,t)
S_t(x_2)\exp[-S_B(t)-S_2(t)].\label{eq:PQCD-factorization}
\end{eqnarray}
This factorization framework  has been successfully generalized  to
a number of transition form factors, including different final
states such as light pseudoscalar and vector
meson~\cite{Lu:2002ny,Chen:2002bq}, scalar
mesons~\cite{Wang:2006ria,Li:2008tk}, axial-vector
mesons~\cite{Wang:2007an,Li:2009tx} and the $D$ meson case in large
recoil region~\cite{Kurimoto:2002sb,Li:2008ts}.

\subsection{$B\to T$ form factors}

%

In analogy with $B\to V$ form factors, we parameterize the $B\to T$
form factors as
 \begin{eqnarray}
  \langle T(P_2,\epsilon)|\bar q\gamma^{\mu}b|\overline B(P_B)\rangle
   &=&-\frac{2V(q^2)}{m_B+m_T}\epsilon^{\mu\nu\rho\sigma} \epsilon^*_{T\nu}  P_{B\rho}P_{2\sigma}, \nonumber\\
  \langle T(P_2,\epsilon)|\bar q\gamma^{\mu}\gamma_5 b|\overline
  B(P_B)\rangle
   &=&2im_T A_0(q^2)\frac{\epsilon^*_{T } \cdot  q }{ q^2}q^{\mu}
    +i(m_B+m_T)A_1(q^2)\left[ \epsilon^*_{T\mu }
    -\frac{\epsilon^*_{T } \cdot  q }{q^2}q^{\mu} \right] \nonumber\\
    &&-iA_2(q^2)\frac{\epsilon^*_{T } \cdot  q }{  m_B+m_T }
     \left[ P^{\mu}-\frac{m_B^2-m_T^2}{q^2}q^{\mu} \right],\nonumber\\
  \langle T(P_2,\epsilon)|\bar q\sigma^{\mu\nu}q_{\nu}b|\overline
  B(P_B)\rangle
   &=&-2iT_1(q^2)\epsilon^{\mu\nu\rho\sigma} \epsilon^*_{T\nu} P_{B\rho}P_{2\sigma}, \nonumber\\
  \langle T(P_2,\epsilon)|\bar q\sigma^{\mu\nu}\gamma_5q_{\nu}b|\overline
  B(P_B)\rangle
   &=&T_2(q^2)\left[(m_B^2-m_T^2) \epsilon^*_{T\mu }
       - {\epsilon^*_{T } \cdot  q }  P^{\mu} \right] +T_3(q^2) {\epsilon^*_{T } \cdot  q }\left[
       q^{\mu}-\frac{q^2}{m_B^2-m_T^2}P^{\mu}\right],\label{eq:BtoTformfactors-definition}
 \end{eqnarray}
where $q=P_B-P_2, P=P_B+P_2$. Similar with the $B\to V$ form
factors, we also have the relation
$2m_TA_0(0)=(m_B+m_T)A_1(0)-(m_B-m_T)A_2(0)$ for tensor mesons in
order to smear the pole at $q^2=0$. In the above definitions the
flavor factor, for instance $1/\sqrt 2$ for the isosinglet meson
with the component $\frac{1}{\sqrt 2}(\bar uu+\bar dd)$, has not
been explicitly specified but  will be taken into account in the
following numerical analysis. The parametrization of $B\to T$ form
factors is analogous to the $B\to V$ case except that the $\epsilon$
is replaced by $\epsilon_T$. In the literature, the $B\to T$ form
factors have been previously defined in an alternative
form~\cite{Isgur:1988gb}
\begin{eqnarray}
 \langle T(P_2,\vp)|V_\mu|\overline B (P_B)\rangle
 &=&-h(q^2)\epsilon_{\mu\nu\alpha\beta}\vp^{\pp*\nu\lambda}P_\lambda
 P^\alpha q^\beta,\nonumber\\
 \langle T(P_2,\vp)|A_\mu|\overline B(P_B)\rangle
 &=&-i\left\{k(q^2)\vp^{\pp*}_{\mu\nu}P^\nu +\vp^{\pp*}_{\alpha\beta}P^\alpha P^\beta[P_\mu b_+(q^2) +q_\mu b_-(q^2)]
 \right\},\label{eq:B TO T-OLD}
\end{eqnarray}
where the two sets of form factors are related via
\begin{eqnarray}
 V&=&-m_B(m_{B}+m_T)h(q^2),\;\;\;\; A_1=-\frac{m_Bk(q^2)}{m_{B}+m_T},\;\;\;
  A_2= m_B(m_{B}+m_T)b_+(q^2),\nonumber\\
 A_0(q^2)&=&\frac{m_{B}+m_T}{2m_T} A_1(q^2)-\frac{m_{B}-m_T}{2m_T}
 A_2(q^2)-\frac{m_Bq^2}{2m_T}b_-(q^2).\label{eq:B TO T1-relation}
\end{eqnarray}

In the PQCD approach, the factorization formulae of $B\to T$ form
factors can be obtained through a straightforward evaluation of the
hard kernels shown in Eq.~\eqref{eq:PQCD-factorization}. But the
correspondence between a vector meson and a tensor meson allows us
to get these formulas in a comparative way. As we have shown in the
above, both LCDAs of a tensor meson and the $B\to T$ form factors
are in conjunction with the quantities involving a vector meson and
explicitly we have
\begin{eqnarray}
 \phi_V^{(i)}\leftrightarrow \phi_T^{(i)},\;\;\;
 F^{B\to T}&\leftrightarrow& F^{B\to
 V},
\end{eqnarray}
where $\phi_{V,T}^{(i)}$ and  $F$ denotes any generic LCDA and $B\to
(T,V)$ form factor, respectively. The only difference is that the
polarization vector $\epsilon$ is replaced by $\epsilon_\bullet$ in
the LCDAs but by $\epsilon_T$ in the transition form factors.  As a
consequence the factorization formulas for the $B\to T$ form factors
are derived as
\begin{eqnarray}
 F^{B\to T}(\phi_T^{(i)})&=& \frac{\epsilon_\bullet}{\epsilon_T} F^{B\to
 V}(\phi_V^{(i)}) =\frac{2 m_B m_T}{m_B^2-q^2} F^{B\to
 V}(\phi_V^{(i)}).\label{eq:comparison}
\end{eqnarray}
As for the expressions of the $B\to V$ form factors, please see
Refs.~\cite{Lu:2002ny,Chen:2002bq} and also  our recent update in
Refs.~\cite{Wang:2007an,Li:2009tx}.

%
 
Form factors in the large recoiling region can be directly
calculated since the exchanged gluon is hard enough so that the
perturbation theory works well. In order to extrapolate the form
factors to the whole kinematic region, we usually use the results
obtained in the region $0< q^2<10\rm{GeV}^2$ and recast the form
factors by adopting certain parametrization of the
$q^2$-distribution. Unlike the other nonperturbative approaches like the QCD sum rules where the analytic properties can be used to constrain the pole structure of the form factors, the PQCD approach is mainly established on the perturbative property of the form factors (i.e. factorization) and in this approach one has to assume the parametrization form in a phenomenological way. 
 In the literature, the popular forms for $B\to P$ and
$B\to V$ form factors (P, V denotes 
a light pseudoscalar meson and a vector meson respectively) include pole form, dipole form and exponential form, and the BK parametrization~\cite{Becirevic:1999kt}. In the small $q^2$ region,
these forms do not differ too much as all of them have similar forms by making use of the expansion of $q^2/m_B^2$. Unfortunately the differences increase with the increase of $q^2$. The limited knowledge of the form factors in the large $q^2$ region will inevitably introduce sizable uncertainties.  However as a first step to proceed, it is helpful to investigate these form factors by employing one commonly-adopted form.  The dipole form has been adopted in the previous PQCD studies~\cite{Li:2008tk,Wang:2007an,Li:2009tx} 
\begin{eqnarray}
 F(q^2)&=&\frac{F(0)}{1-a(q^2/m_B^2)+b(q^2/m_B^2)^2}
\end{eqnarray}
and this parametrization works well. 
In contrast, the $B\to T$ form factors receive additional
$q^2$-dependence as can be seen from the factorization formulas in
Eq.~\eqref{eq:comparison}. In this case the following modified form
is more appropriate for the $q^2$-distribution of $B\to T$ form
factors
\begin{eqnarray}
 F(q^2)&=&\frac{F(0)}{(1-q^2/m_B^2)(1-a(q^2/m_B^2)+b(q^2/m_B^2)^2)},\label{eq:fit-B-T}
\end{eqnarray}
and we shall use this form in our fitting procedure.

Numerical results for the form factors at maximally recoil point and
the two fitted parameters $a,b$ are collected in
table~\ref{Tab:formfactorsBtoTbeforemixing}. The first type of
errors comes from decay constants and shape parameter $\omega_b$ of
$B$ meson; while the second one is from factorization scales (from
$0.75t$ to $1.25t$, not changing the transverse part $1/b_i$), the
threshold resummation parameter $c=0.4\pm0.1$ and $\Lambda_{\rm
QCD}=(0.25\pm0.05)\rm{GeV}$. The hadron masses are taken from
particle data group~\cite{Amsler:2008zz}
\begin{eqnarray}
 m_{a_2}=1.3183{\rm GeV},\;\; m_{K_2^*}=1.43 {\rm GeV},\;\; m_{f_2(1270)}=1.2751{\rm
 GeV},\;\; m_{f_2'(1525)}=1.525 {\rm GeV}.
\end{eqnarray}

\begin{table}
\caption{$B\to T$ form factors. $a,b$ are the parameters of the form
factors in the parametrization shown in Eq.~\eqref{eq:fit-B-T}. The
two kinds of errors are from: decay constants of $B$ meson and shape
parameter $\omega_b$; $\Lambda_{\rm{QCD}}$, the scales $t$s and the
threshold resummation parameter $c$. }
 \label{Tab:formfactorsBtoTbeforemixing}
 \begin{center}
 \begin{tabular}{ c c c ccc c c}
\hline \hline
 $F$       & $F(0)$  & $a$ &$b$       & $F$       & $F(0)$  & $a$ &$b$                \\
 \hline
\hline
  $V^{B a_2}$       & $0.18_{-0.03-0.03}^{+0.04+0.04}$
                    & $1.70_{-0.01-0.05}^{+0.01+0.06}$
                    & $0.63_{-0.01-0.04}^{+0.03+0.09}$
  &$V^{B f_2(1270)}$& $0.12_{-0.02-0.02}^{+0.02+0.02}$
                    & $1.68_{-0.00-0.05}^{+0.02+0.06}$
                    & $0.62_{-0.00-0.07}^{+0.05+0.10}$  \\ \hline
  $A_0^{B a_2}$     & $0.18_{-0.03-0.03}^{+0.04+0.04}$
                    & $1.74_{-0.05-0.07}^{+0.00+0.06}$
                    & $0.71_{-0.13-0.13}^{+0.00+0.07}$
  &$A_0^{Bf_2(1270)}$&$0.13_{-0.02-0.02}^{+0.03+0.03}$
                    & $1.74_{-0.02-0.06}^{+0.01+0.05}$
                    & $0.69_{-0.05-0.10}^{+0.04+0.06}$ \\ \hline
  $A_1^{B a_2}$     & $0.11_{-0.02-0.02}^{+0.02+0.02}$
                    & $0.74_{-0.01-0.03}^{+0.02+0.04}$
                    & $-0.11_{-0.03-0.02}^{+0.04+0.03}$
  &$A_1^{B f_2(1270)}$&$0.08_{-0.01-0.01}^{+0.02+0.01}$
                    & $0.73_{-0.03-0.04}^{+0.01+0.05}$
                    & $-0.12_{-0.09-0.00}^{+0.03+0.04}$ \\ \hline
  $A_2^{B a_2}$     & $0.06_{-0.01-0.01}^{+0.01+0.01}$  &$--$  &$--$
  &$A_2^{B f_2(1270)}$&$0.04_{-0.01-0.00}^{+0.01+0.01}$ &$--$  &$--$\\ \hline
  $T_1^{B a_2}$     & $0.15_{-0.03-0.02}^{+0.03+0.03}$
                    & $1.69_{-0.01-0.05}^{+0.00+0.05}$
                    & $0.64_{-0.04-0.06}^{+0.00+0.05}$
  &$T_1^{B f_2(1270)}$ &$0.10_{-0.02-0.01}^{+0.02+0.02}$
                    & $1.67_{-0.01-0.08}^{+0.00+0.05}$
                    & $0.62_{-0.03-0.15}^{+0.00+0.05}$ \\ \hline
 $T_2^{B a_2}$      & $0.15_{-0.03-0.02}^{+0.03+0.03}$
                    & $0.74_{-0.01-0.07}^{+0.01+0.01}$
                    & $-0.11_{-0.01-0.09}^{+0.02+0.00}$
 &$T_2^{B f_2(1270)}$&$0.10_{-0.02-0.01}^{+0.02+0.02}$
                    & $0.72_{-0.04-0.08}^{+0.00+0.03}$
                    & $-0.09_{-0.10-0.11}^{+0.00+0.00}$  \\ \hline
  $T_3^{B a_2}$     & $0.13_{-0.02-0.02}^{+0.03+0.03}$
                    & $1.58_{-0.01-0.05}^{+0.01+0.06}$
                    & $0.52_{-0.04-0.04}^{+0.02+0.05}$
  &$T_3^{B f_2(1270)}$&$0.09_{-0.02-0.01}^{+0.02+0.02}$
                    & $1.56_{-0.00-0.05}^{+0.03+0.08}$
                    & $0.48_{-0.00-0.04}^{+0.08+0.12}$   \\
 \hline\hline
 $V^{B K_{2}^*}$    & $0.21_{-0.04-0.03}^{+0.04+0.05}$
                    & $1.73_{-0.02-0.03}^{+0.02+0.05}$
                    & $0.66_{-0.05-0.01}^{+0.04+0.07}$  \\ \hline
 $A_0^{B K_{2}^*}$  & $0.18_{-0.03-0.03}^{+0.04+0.04}$
                    & $1.70_{-0.02-0.07}^{+0.00+0.05}$
                    & $0.64_{-0.06-0.10}^{+0.00+0.04}$ \\ \hline
 $A_1^{B K_{2}^*}$  & $0.13_{-0.02-0.02}^{+0.03+0.03}$
                    & $0.78_{-0.01-0.04}^{+0.01+0.05}$
                    & $-0.11_{-0.03-0.02}^{+0.02+0.04}$  \\ \hline
 $A_2^{B K_{2}^*}$  & $0.08_{-0.02-0.01}^{+0.02+0.02}$  &$--$  &$--$       \\
 \hline
 $T_1^{B K_{2}^*}$  & $0.17_{-0.03-0.03}^{+0.04+0.04}$
                    & $1.73_{-0.03-0.07}^{+0.00+0.05}$
                    & $0.69_{-0.08-0.11}^{+0.00+0.05}$  \\
 \hline
 $T_2^{B K_{2}^*}$  & $0.17_{-0.03-0.03}^{+0.03+0.04}$
                    & $0.79_{-0.04-0.09}^{+0.00+0.02}$
                    & $-0.06_{-0.10-0.16}^{+0.00+0.00}$    \\
 \hline
 $T_3^{B K_{2}^*}$  & $0.14_{-0.03-0.02}^{+0.03+0.03}$
                    & $1.61_{-0.00-0.04}^{+0.01+0.09}$
                    & $0.52_{-0.01-0.01}^{+0.05+0.15}$   \\
 \hline \hline
  $V^{B_s K_2^*}$   & $0.18_{-0.03-0.03}^{+0.03+0.04}$
                    & $1.73_{-0.00-0.05}^{+0.02+0.05}$
                    & $0.67_{-0.00-0.05}^{+0.05+0.06}$
  &$V^{B_s f_2'(1525)}$& $0.20_{-0.03-0.03}^{+0.04+0.05}$
                    & $1.75_{-0.00-0.03}^{+0.02+0.05}$
                    & $0.69_{-0.01-0.01}^{+0.05+0.08}$  \\
 \hline
  $A_0^{B_s K_2^*}$ & $0.15_{-0.02-0.02}^{+0.03+0.03}$
                    & $1.70_{-0.01-0.05}^{+0.00+0.03}$
                    & $0.65_{-0.03-0.04}^{+0.01+0.00}$
  &$A_0^{B_s f_2'(1525)}$ &$0.16_{-0.02-0.02}^{+0.03+0.03}$
                    & $1.69_{-0.01-0.03}^{+0.00+0.04}$
                    & $0.64_{-0.04-0.02}^{+0.00+0.01}$\\
 \hline
  $A_1^{B_s K_2^*}$ & $0.11_{-0.02-0.02}^{+0.02+0.02}$
                    & $0.79_{-0.01-0.03}^{+0.02+0.03}$
                    & $-0.10_{-0.03-0.02}^{+0.07+0.06}$
  &$A_1^{B_s f_2'(1525)}$&$0.12_{-0.02-0.02}^{+0.02+0.03}$
                    & $0.80_{-0.00-0.03}^{+0.02+0.07}$
                    & $-0.11_{-0.00-0.00}^{+0.05+0.09}$ \\
 \hline
 $A_2^{B_s K_2^*}$  & $0.07_{-0.01-0.01}^{+0.01+0.02}$  &$--$  &$--$
 &$A_2^{B_s f_2'(1525)}$&$0.09_{-0.01-0.01}^{+0.02+0.02}$ &$--$  &$--$\\
 \hline
 $T_1^{B_s K_2^*}$  & $0.15_{-0.02-0.02}^{+0.03+0.03}$
                    & $1.73_{-0.01-0.06}^{+0.00+0.04}$
                    & $0.69_{-0.03-0.11}^{+0.00+0.04}$
 &$T_1^{B_s f_2'(1525)}$&$0.16_{-0.03-0.02}^{+0.03+0.04}$
                    & $1.75_{-0.00-0.05}^{+0.01+0.05}$
                    & $0.71_{-0.01-0.08}^{+0.03+0.06}$  \\
 \hline
 $T_2^{B_s K_2^*}$  & $0.15_{-0.02-0.02}^{+0.03+0.03}$
                    & $0.80_{-0.03-0.08}^{+0.00+0.02}$
                    & $-0.06_{-0.09-0.13}^{+0.00+0.00}$
 &$T_2^{B_s f_2'(1525)}$&$0.16_{-0.03-0.02}^{+0.03+0.04}$
                    & $0.82_{-0.04-0.06}^{+0.00+0.04}$
                    & $-0.08_{-0.09-0.08}^{+0.00+0.03}$  \\
 \hline
  $T_3^{B_s K_2^*}$ & $0.12_{-0.02-0.02}^{+0.02+0.03}$
                    & $1.61_{-0.00-0.04}^{+0.03+0.08}$
                    & $0.52_{-0.01-0.00}^{+0.08+0.14}$
  &$T_3^{B_s f_2'(1525)}$&$0.13_{-0.02-0.02}^{+0.03+0.03}$
                    & $1.64_{-0.00-0.06}^{+0.02+0.06}$
                    & $0.57_{-0.01-0.09}^{+0.04+0.05}$   \\
 \hline\hline
 \end{tabular}
 \end{center}
 \end{table}

\begin{figure}
\includegraphics[scale=0.7]{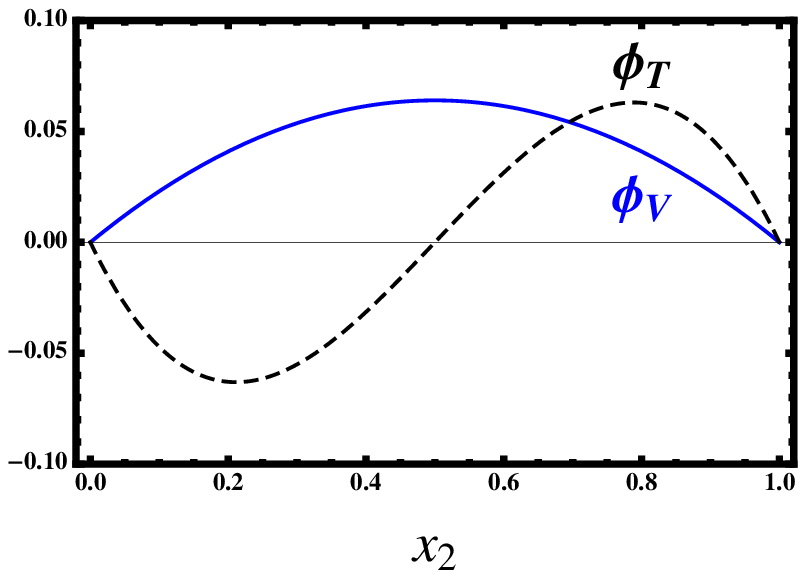}
\includegraphics[scale=0.7]{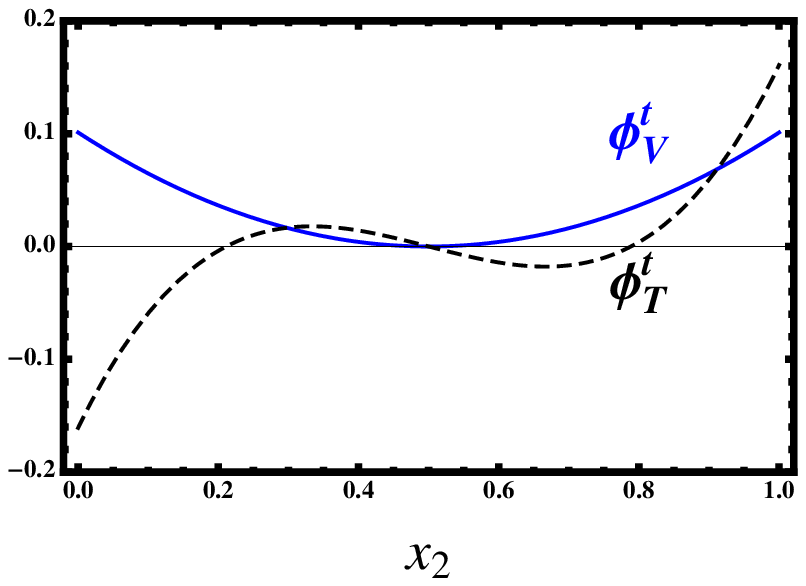}
\includegraphics[scale=0.7]{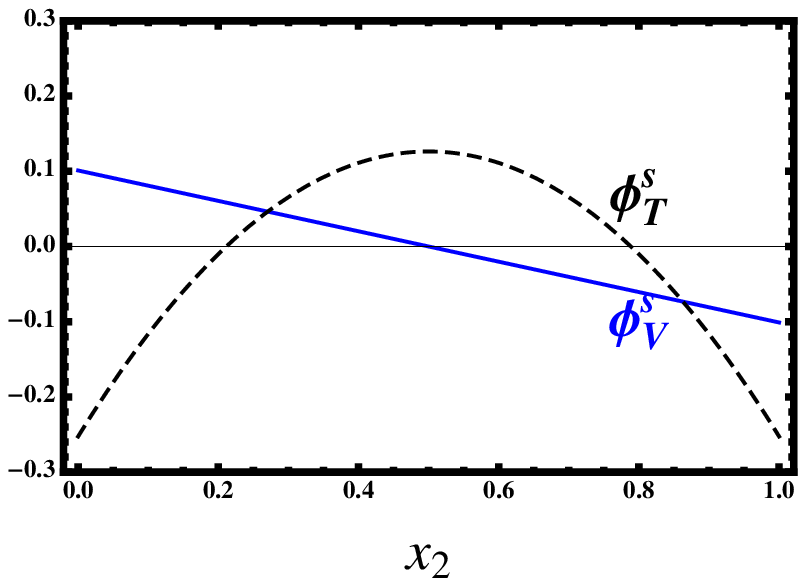}
\includegraphics[scale=0.7]{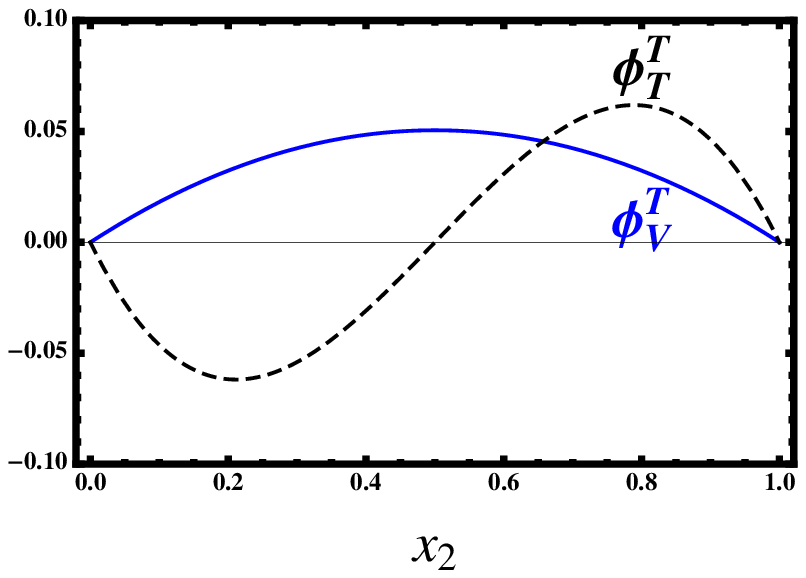}
\includegraphics[scale=0.7]{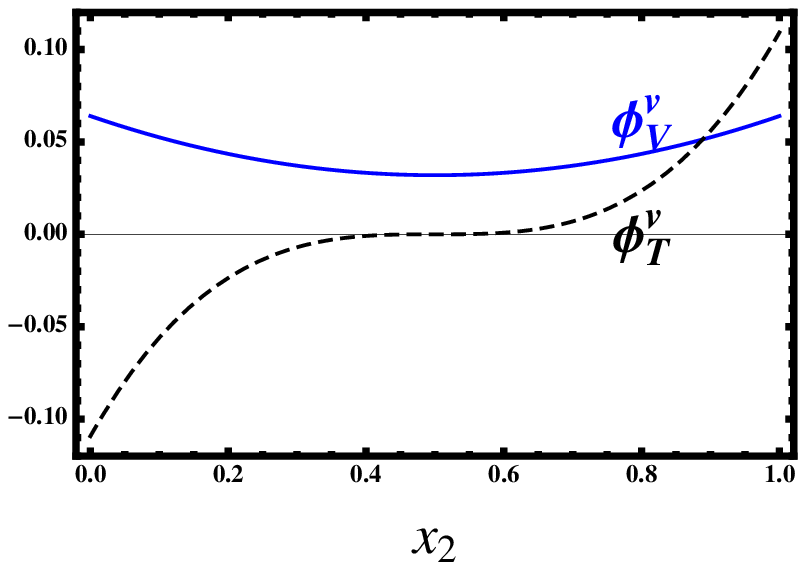}
\includegraphics[scale=0.7]{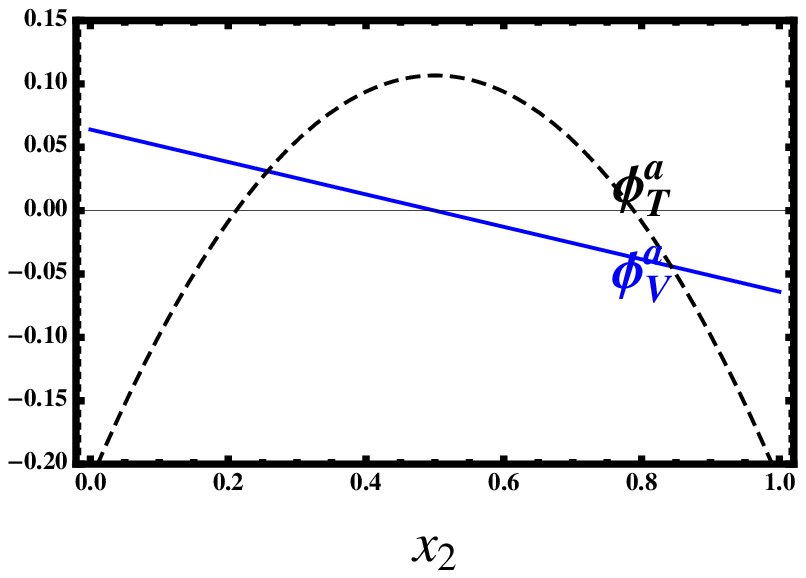}
\caption{LCDAs of the vector meson $\rho$ (solid lines) and its
tensor counterpart $a_2$ (dashed lines). The asymptotic forms are
adopted for $\rho,a_2$ meson LCDAs. }\label{fig:LCDA-comparison}
\end{figure}

\begin{figure}
\includegraphics[scale=0.7]{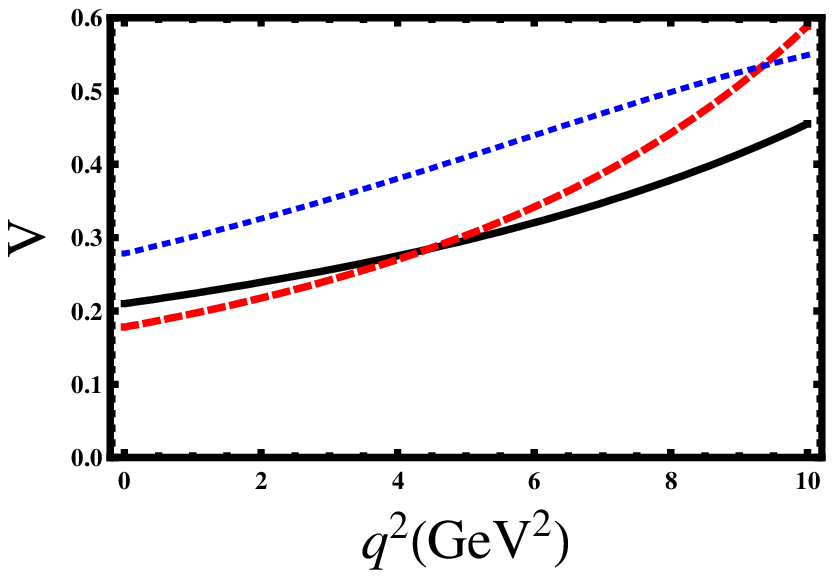}
\includegraphics[scale=0.7]{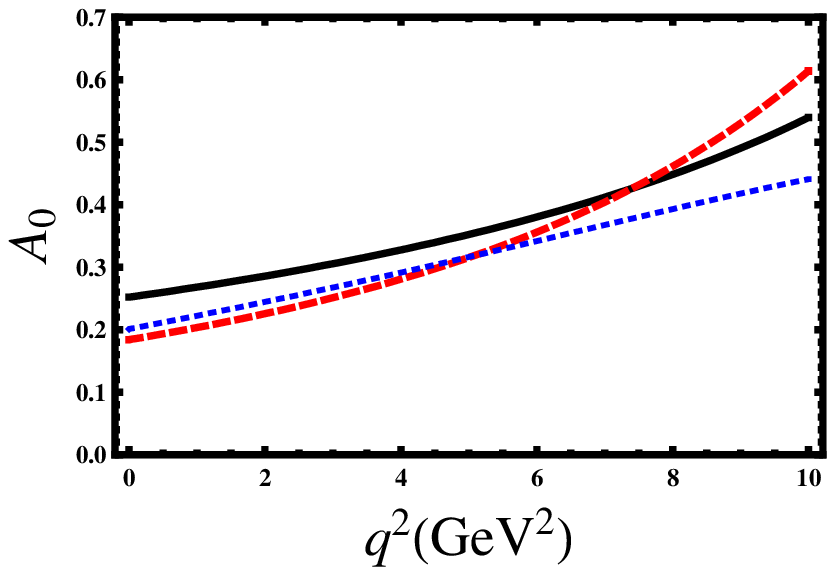}
\includegraphics[scale=0.7]{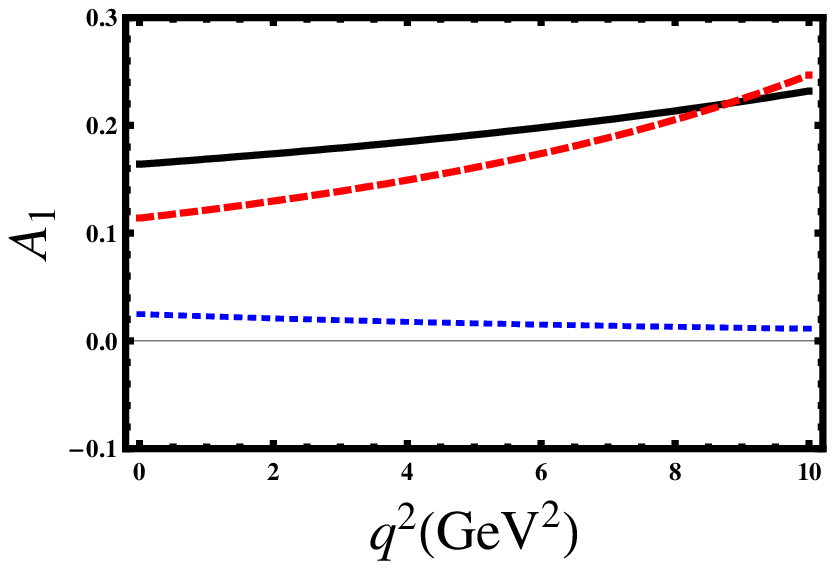}
\includegraphics[scale=0.7]{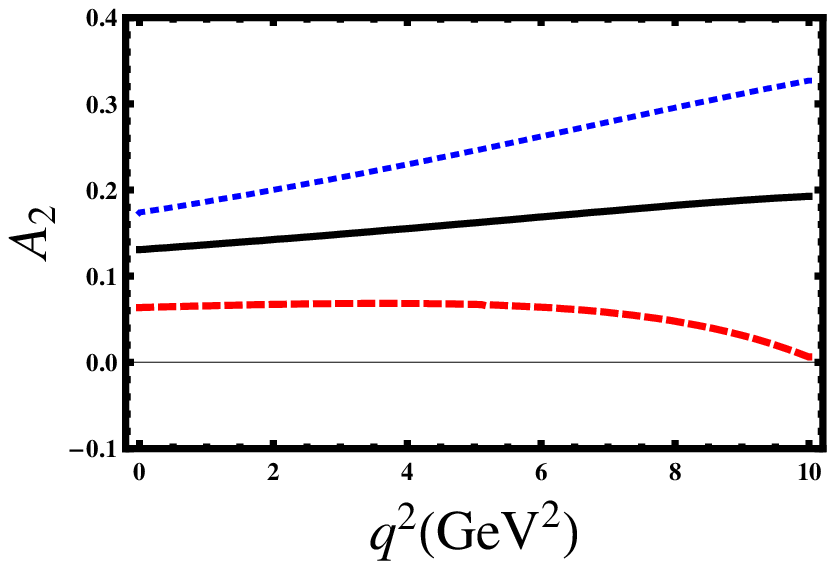}
\includegraphics[scale=0.7]{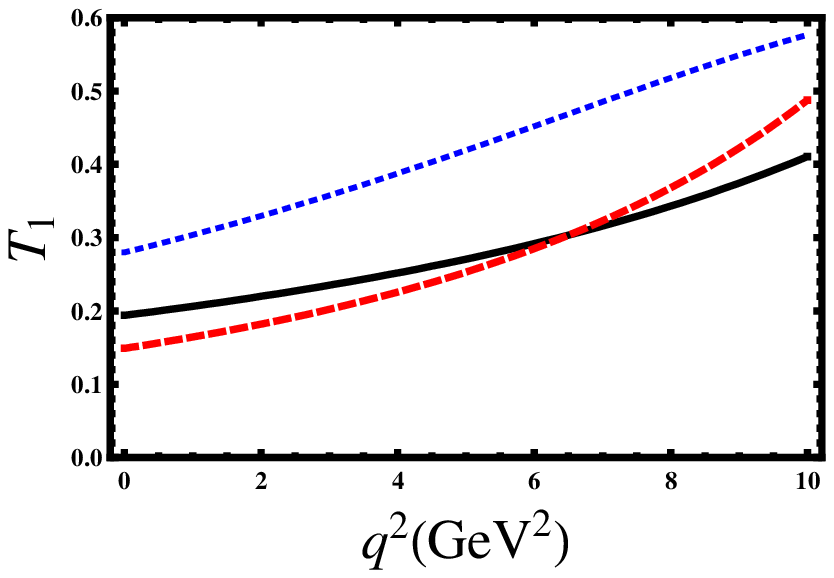}
\includegraphics[scale=0.7]{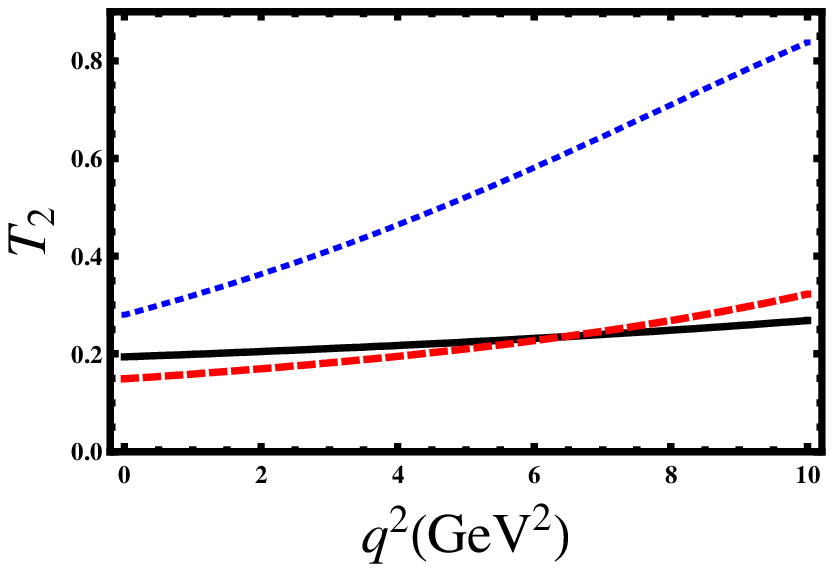}
\includegraphics[scale=0.7]{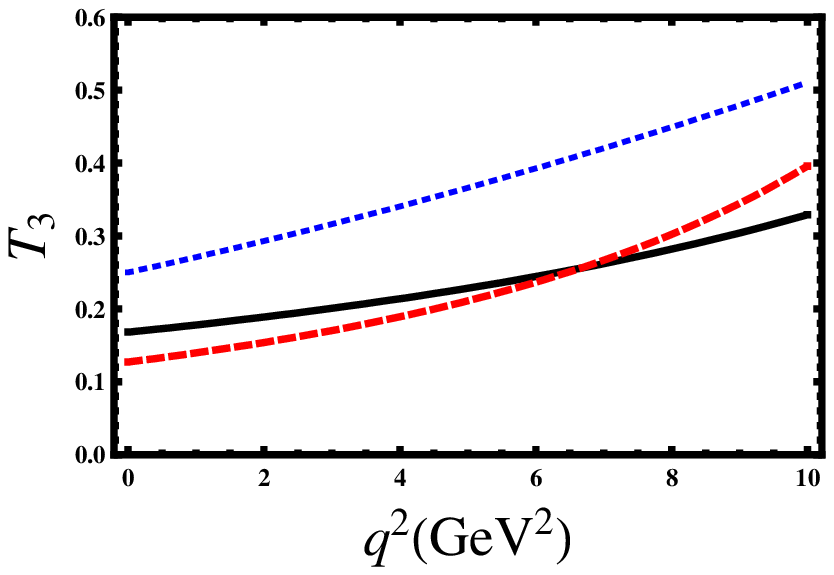}
\caption{Transition form factors as functions of $q^2$. Solid
(black) and dashed (red) lines correspond to our results of the
$B\to\rho$ and $B\to a_2$ channel, respectively. Dotted (blue) lines
denote the results in the covariant LFQM, with  $V,A_0,A_1,A_2$ for
the $B\to a_2$ process and $T_{1,2,3}$ for the $B\to K_2^*$
transition. A minus sign has been added to the LFQM results for
$V,A_1,A_2,T_3$ so that they have the same sign with our results.
}\label{fig:q2-dependence}
\end{figure}

A number of remarks on these results are given in order.
\begin{enumerate}

\item With terms suppressed by $r_2^2$
neglected, $A_2(q^2)$ can be expressed  as a linear combination of
$A_0$ and $A_1$~\cite{Li:2009tx}
\begin{eqnarray}
  A_2(q^2)=\frac{1+r_2}{1-q^2/m_B^2}\left[(1+r_2)A_1(q^2)-2r_2
  A_0(q^2)\right].
\end{eqnarray}
We will use this relation for $A_2(q^2)$ in the whole kinematic
region instead of a direct fitting.

\item The $B\to f_2(1270)$ form factors are smaller than the other channels due to the
factor $1/\sqrt2$ in the flavor wave function of $f_2(1270)$. The
smaller transverse decay constants of $K_2^*$ and $f_2'(1525)$ have
a tendency to suppress the transition amplitudes. But
their larger  
masses give an enhancement, since both contributions from the twist-3 LCDAs and the correspondence relation in Eq.~(\ref{eq:comparison}) are proportional to the hadron mass.

\item The parameters $a$ in most transition form factors are roughly $1.7$,
but they are around $0.7$ for $A_1(q^2)$ and $T_2(q^2)$. Analogously
the parameter $b$ is close to $0.6$ with the exception for
$A_1(q^2)$ and $T_2(q^2)$ as it is approaching $0$.  The vanishing
$b$ implies that the dipole behavior in these two form factors is
reduced into the monopole form.

\item In our computation, the asymptotic forms for the LCDAs have been adopted. 
The twist-2 LCDAs $\phi_{||,\perp}$ can be expanded into Gegenbauer polynomials $C_{n}^{3/2}(2x_2-1)$ (with $x_2$ being the momentum fraction of the quark in the meson) and the twist-3 LCDAs will be expressed in terms of twist-2 ones 
through the use of equation of motion~\cite{Cheng:2010hn}:
\begin{eqnarray}
 g_\perp^{(v)}(x_2)&=& \int_0^{x_2} dv \frac{\phi_{||}(v)}{1-v}+ \int_{x_2}^1 dv \frac{\phi_{||}(v)}{v},\nonumber\\
  g_{\perp}^{(a)}(x_2)&=& 4(1-x_2) \int_0^{x_2} dv \frac{\phi_{||}(v)}{1-v}
 +4x_2\int_{x_2}^1 dv \frac{\phi_{||}(v)}{v},\nonumber\\
 h_{||}^{(t)}(x_2)&=& \frac{3}{2} (2x_2-1) \left[\int_0^{x_2} dv\frac{\phi_{\perp}(v)}{1-v}-\int_{x_2}^1 dv\frac{\phi_{\perp}(v)}{v}\right],\nonumber\\
 h_{||}^{(s)}(x_2)&=&3(1-x_2) \int_0^{x_2} dv \frac{\phi_{\perp}(v)}{1-v}
 +3x_2\int_{x_2}^1 dv \frac{\phi_{\perp}(v)}{v}.
\end{eqnarray}
Taking into account the contributions from the next non-zero Gegenbauer moment besides the asymptotic form, i.e. $a_3$,  we find
\begin{eqnarray}
 A_0^{Ba_2}(0)=0.18\pm0.07 a_3,\;\;\; T_1^{Ba_2}(0)= 0.15\pm 0.057 a_3. 
\end{eqnarray}
In the case of $\pi$ and $\rho$ meson, the first non-zero Gegenbauer moment  is around (0.2-0.3)~\cite{Ball:2006wn}. 
If it were the similar for the tensor meson,
we can see that the form factors will be changed by roughly $10\% -20\%$.

\item Since the $B\to T$ form factors are obtained from the $B\to V$
ones, it is meaningful to analyze these two sets of form factors in
a comparative way.  It is worth comparing
their distribution amplitudes. 
The six LCDAs are functions of $x_2$, with $x_2$ being the momentum fraction of the quark in the light meson. 
Taking $\rho$ and $a_2$ mesons as an example, these LCDAs
are depicted in Fig.~\ref{fig:LCDA-comparison},  where the solid
(dashed) lines denote the LCDAs for $\rho$ ($a_2$) meson. For $\rho$
meson LCDAs, the asymptotic form has been used. From this figure, we
can see that although the two sets of LCDAs are different in the
small-momentum-fraction region $x_2<0.5$, they have similar shapes
when $x_2>0.6$. The large-momentum-fraction region, $0.6<x_2<1$~\footnote{
The dominant region in the PQCD approach can be obtained by the power counting in this approach which has been established in Ref.~\cite{Chen:2001pr}.
The typical momentum  of the spectator (with the momentum fraction $1-x_2$) is of the order $\Lambda/m_B<0.3$ with the vary of the hadronization scale 
$\Lambda$.  This means that the dominant contribution lies in the region of $0.7<x_2$. 
This conclusion can also be drawn in a simple way. 
PQCD is based on the hard scattering picture, in which the endpoint region $x_2\sim 0,1$ is suppressed by the Sudakov factor. In the Feynman diagrams given in Fig.1,
if the momentum of the spectator (with momentum fraction $1-x_2$) is getting larger, the gluon and the quark propagators will have larger virtualities. For instance, 
they are $p_b^2-m_b^2= x_2\eta m_B^2- k_{1\perp}^2$ and $p_g^2=x_1x_2\eta m_B^2 -(k_{1\perp}-k_{2\perp})^2$ with the transverse component $k_{1\perp,2\perp}$  of the order $\Lambda$.   
Therefore the region $x_2>0.5$ is more important  compared with the region of $x_2<0.5$ and thus in our analysis the region of $x_2>0.6$ is chosen. }
dominates in the PQCD approach. As one important consequence, the
$B\to T$ and $B\to V$ form factors will have several similar
properties. For instance the two kinds of form factors will have the
same signs and their $q^2$-dependence parameters will also be close.


\item As functions of $q^2$, the $B\to T$ form factors are
expected to be sharper than the $B\to V$ form factors, since the
former ones contain one more pole structure in the
$q^2$-distribution. To illustrate this situation, in
Fig.~\ref{fig:q2-dependence} we show the $B\to \rho$ (dashed lines)
and $B\to a_2$ form factors (solid lines) in the region of $0<q^2<10
{\rm GeV}^2$, where the PQCD results for the $B\to \rho$ form
factors are taken from our recent update in Ref.~\cite{Li:2009tx}.
We also quote them in table~\ref{Tab:Btorhoformfactor}, but only the
central values are shown for the $q^2$-dependence parameters $a,b$.
The ratio of the $B\to \rho$ and $B\to a_2$ form factors is 0.73 for
$A_0$ and 0.77 for $T_1$, respectively.

\begin{table}
\caption{$B\to \rho$ form factors in the PQCD
approach~\cite{Li:2009tx}  }
\begin{center}
\begin{tabular}{ccccccc}
\hline \hline
     & $F$ & $F(0)$ &$a$ &$b$ \\\hline
 & $V$    &$0.21_{-0.04-0.02}^{+0.05+0.03}$      & 1.75 & 0.69   \\
       & $A_0$  &$0.25_{-0.05-0.03}^{+0.06+0.04}$ & 1.69 & 0.57       \\
       & $A_1$  &$0.16_{-0.03-0.02}^{+0.04+0.02}$ & 0.77 & $-0.13$   \\
      & $A_2$  &$0.13_{-0.03-0.01}^{+0.03+0.02}$  &--- &---   \\
      & $T_1$  &$0.19_{-0.04-0.02}^{+0.04+0.03}$  & 1.69 &0.61 \\
      & $T_2$  &$0.19_{-0.04-0.02}^{+0.04+0.03}$ &0.73 &$-0.12$ \\
      & $T_3$  &$0.17_{-0.03-0.02}^{+0.04+0.02}$  & $1.58$ & 0.50  \\

 \hline \hline
\end{tabular}\label{Tab:Btorhoformfactor}
\end{center}
\end{table}

\item  At the maximally recoiling point with
$q^2=0$, the $B\to\rho$ and $B\to a_2(1320)$ form factors have
different magnitudes. Taking $A_0$ and $T_1$ as an example,  in
table~\ref{Tab:formfactorcomparison} we enumerate distinct
contributions from the three LCDAs. The matching coefficient
$2m_Tm_B/(m_B^2-q^2)$ between the two sets of form factors is
roughly $1/2$ at $q^2=0$ and in this case the $B\to T$ transition is
expected to be smaller. It is also confirmed by the numerical
results in table~\ref{Tab:formfactorcomparison}, where its twist-2
contribution is only one half of the $B\to V$ case. On the contrary
this does not occur for the twist-3 LCDAs, as the larger tensor
meson mass has compensated the suppression: $m_{a_2}\sim 2m_{\rho}$.

\begin{table}[tb]
\caption{Different contributions to form factors $A_0$ and $T_1$ for
$B\to \rho$ and $B\to a_2(1320)$.}
 \label{Tab:formfactorcomparison}
\begin{center}
 \begin{tabular}{cccccc}
  \hline\hline
        $A_0$    &  $B\to\rho$  & $B\to a_2(1320)$
          \\  \hline
   $\phi$
                                                          &$0.108$ &
                                                          $0.050$
                                                          \\\hline
   $\phi^s$
                                                          &$0.103$&
                                                          $0.088$
                                                          \\ \hline
   $\phi^t$
                                                          &$0.040$
                                                          & $0.046$
                                                           \\
 \hline
 total                                                    &$0.251$
                                                          & $0.184$
                                                          \\%
 \hline\hline
        $T_1$    &  $B\to\rho$ & $B\to a_2(1320)$
          \\  \hline
   $\phi^T$
                                                          &$0.085$
                                                          &$0.049$
                                                          \\\hline
   $\phi^a$
                                                          &$0.047$&
                                                          $0.046$
                                                          \\ \hline
   $\phi^v$
                                                          &$0.063$
                                                          &$0.054$
                                                           \\
 \hline
 total                                                    &$0.194$
                                                          &$0.150$
                                                          \\
 \hline\hline
\end{tabular}
\end{center}
\end{table}

\end{enumerate}

In the literature, the $B\to T$ form factors have been explored in
the ISGW model~\cite{Isgur:1988gb}, its improved form ISGW II
model~\cite{Scora:1995ty,Kim:2002rx,Sharma:2010yx,Cheng:2010sn} and
other relativistic quark models for instance the covariant
light-front quark model
(LFQM)~\cite{Cheng:2003sm,Cheng:2004yj,Cheng:2009ms}. The form
factor $T_1$ for $B\to K_2^*$ is also estimated in the technique of
QCD sum rules (QCDSR)~\cite{Safir:2001cd}, relativistic quark
model~\cite{Ebert:2001en} and heavy quark
symmetry~\cite{Veseli:1995bt}.  We collect the results using these
approaches
\cite{Sharma:2010yx,Cheng:2010sn,Cheng:2003sm,Cheng:2004yj,Cheng:2009ms,Safir:2001cd}
in table~\ref{Tab:BtoVformfactor} for the convenience of a
comparison, where their results have been converted to the new form
factors defined in Eq.~\eqref{eq:BtoTformfactors-definition} through
the relations in Eq.~\eqref{eq:B TO T1-relation}. Our PQCD results,
all uncertainties added in quadrature, are also shown in
table~\ref{Tab:BtoVformfactor}. From this table, we can find many
differences among these theoretical predictions. Results for all
form factors from the ISGW II model possess a different sign with
our results and the magnitudes are typically larger. The two
calculations in the same ISGW II model  are also different, for
instance the prediction in Ref.~\cite{Sharma:2010yx} of $A_2$ for
$B\to K_2^*$ is about twice as large as the one in
Ref.~\cite{Cheng:2010sn}. The estimate in the
QCDSR~\cite{Safir:2001cd} is consistent with our result.

Results in the covariant LFQM are different with ours in several
aspects. Firstly, for $A_0$ and $T_{1,2}$~\footnote{The form factors
$T_{1,2,3}$ in this work correspond to the $U_{1,2,3}$ in
Ref.~\cite{Cheng:2009ms}.}, the LFQM predicts the same sign with our
results but the remanent results have negative signs. Secondly,
their predictions, except for $A_1$, are much larger than ours in
magnitude. Moreover the $q^2$-distribution is also different.  In
Fig.~\ref{fig:q2-dependence}, we show the LFQM results (dotted
lines) in the region of $0<q^2<10 {\rm GeV}^2$, with $V,A_{0,1,2}$
for the $B\to a_2$ process~\cite{Cheng:2003sm} but $T_{1,2,3}$ for
the $B\to K_2^*$ transition~\cite{Cheng:2009ms}. A minus sign has
been added to $V,A_1,A_2,T_3$ so that they have the same sign with
our results. From this figure, we can find that the differences for
$A_1,A_2,T_2$ between their results and ours get larger as $q^2$
grows. In particular, the $T_2$ grows faster than $T_1$ with the
increase of $q^2$ in the LFQM but it is reverse in our results. In
the covariant LFQM the meson-quark-antiquark coupling vertex for a
tensor meson contains
$\epsilon_{\mu\nu}\frac{p_1^{\prime\nu}-p_2^\nu}{2}
\sqrt{\frac{2}{\beta^{\prime2}}}$, which corresponds to
$\epsilon_\mu$ in the case of a vector meson.  $p_1'(p_2)$ denotes
the momentum of the quark and antiquark in the final meson. The
$\beta'$, of the order $\Lambda_{\rm QCD}$, is the shape parameter
which characterizes the momentum distribution inside the tensor
meson. It is hard to deduce the relative signs from this structure
since (1) apart from the longitudinal momentum in $p_1',p_2$, the
transverse part might also contribute; (2) it involves the zero-mode
terms which are essential for the maintenance of the Lorentz
covariance. In this sense the relation between a vector meson and
its tensor counterpart is not as simple as the one in the PQCD
approach, where $\epsilon$ is replaced by $\epsilon_\bullet$. These
different results can be discriminated in the future when enough
data is available.

\begin{table}
\caption{$B\to T$ form factors at maximally recoil, i.e. $q^2=0$.
Theoretical results in the ISGW II model~\cite{Sharma:2010yx}, the
covariant light-front quark model~\cite{Cheng:2003sm,Cheng:2009ms}
and the QCD sum rules~\cite{Safir:2001cd} are also collected for a
comparison.  Results in the parentheses are from
Ref.~\cite{Cheng:2010sn}.}
\begin{center}
\begin{tabular}{cccccccc}
\hline \hline
         &       & $B\to a_2$ &$B\to K_2^*$     &$B\to f_2$ &$B_s\to K_2^*$  &$B_s\to f_2'$ \\
 \hline
 ISGW II~\cite{Sharma:2010yx}(\cite{Cheng:2010sn})  &$V$ &  &---$(-0.57)$           &             &              &        \\
    & $A_0$                 &$-0.18$      &$-0.17(-0.25)$ &  $-0.08$   & $-0.27$ & $-0.26$       \\
     & $A_1$             &$-0.35$      &$-0.38(-0.23)$         & $-0.24$    & $-0.39$     & $-0.45$    \\
      & $A_2$           &$-0.45$      &$-0.53(-0.21)$        &  $-0.34$   &  $-0.47$      & $-0.59$     \\
\hline  \hline
LFQM~\cite{Cheng:2003sm,Cheng:2009ms}&$V$ &$-0.28$      &$-0.28$        &             &              &        \\
      & $A_0$            &$0.20$      &$0.26$         &             &              &        \\
     & $A_1$             &$-0.025$      &$-0.012$         &             &              &        \\
      & $A_2$           &$-0.17$      &$-0.21$        &             &              &        \\
      & $T_1=T_2$       &
             &$0.28$         &             &              & $0.28$        \\
      & $T_3$           &       &$-0.25$        &             &              & $-0.18$       \\
\hline QCDSR~\cite{Safir:2001cd}
      & $T_1$       &
             &$0.19\pm0.04$         &             &              &
             \\
\hline
 This work  & $V$    & $0.18_{-0.04}^{+0.05}$    & $0.21_{-0.05}^{+0.06}$
     &   $   0.12_{-0.03}^{+0.03}$    & $0.18_{-0.04}^{+0.05}$&
     $0.20_{-0.04}^{+0.06}$ \\
       & $A_0$  &$   0.18
            _{  -0.04         }
            ^{+   0.06        }
 $     & $   0.18
            _{  -0.04         }
            ^{+   0.05        }
 $    &$   0.13
            _{  -0.03         }
            ^{+   0.04        }
 $      &$   0.15
            _{  -0.03         }
            ^{+   0.04        }
 $&$   0.16
            _{  -0.03         }
            ^{+   0.04        }
 $\\
       & $A_1$  &$   0.11
            _{  -0.03         }
            ^{+   0.03        }
 $    &$   0.13
            _{  -0.03         }
            ^{+   0.04        }
 $      &$   0.08
            _{  -0.02         }
            ^{+   0.02        }
 $      &$   0.11
            _{  -0.02         }
            ^{+   0.03        }
 $ &$   0.12
            _{  -0.03         }
            ^{+   0.03        }
 $\\
      & $A_2$  &$   0.06
            _{  -0.01         }
            ^{+   0.02        }
 $     &$   0.08
            _{  -0.02         }
            ^{+   0.03        }
 $       &$   0.04
            _{  -0.01         }
            ^{+   0.01        }
 $      &$   0.07
            _{  -0.02         }
            ^{+   0.02        }
 $& $   0.09
            _{  -0.02         }
            ^{+   0.03        }
 $\\
      & $T_1=T_2$  &$   0.15
            _{  -0.03         }
            ^{+   0.04        }
 $     &$   0.17
            _{  -0.04         }
            ^{+   0.05        }
 $       &$   0.10
            _{  -0.02         }
            ^{+   0.03        }
 $      &$   0.15
            _{  -0.03         }
            ^{+   0.04        }
 $&$   0.16
            _{  -0.04         }
            ^{+   0.05        }
 $\\
      & $T_3$  &$   0.13
            _{  -0.03         }
            ^{+   0.04        }
 $  &$   0.14
            _{  -0.03         }
            ^{+   0.05        }
 $       &$   0.09
            _{  -0.02         }
            ^{+   0.03        }
 $    &$   0.12
            _{  -0.03         }
            ^{+   0.04        }
 $&$   0.13
            _{  -0.03         }
            ^{+   0.04        }
 $\\
 \hline \hline
\end{tabular}\label{Tab:BtoVformfactor}
\end{center}
\end{table}

In the large energy limit, the seven $B\to T$ form factors are
expected to satisfy several nontrivial
relations~\cite{Datta:2007yk,Hatanaka:2009gb} and  all form factors
can be parameterized into two independent functions
$\zeta_\perp(q^2)$ and $\zeta_{||}(q^2)$. In the large recoil
region, we have checked that our results respect these relations.
Moreover, the relative size of these two functions is also of prime
interest but it can not be deduced from the large energy limit
itself. Our results for the $B\to K_2^*$ transition~\footnote{ We
use the definitions of $\zeta_{\perp}$ and $\zeta_{||}$  in
Ref.~\cite{Hatanaka:2009gb}, but our form factors correspond to
theirs with a tilde. },
\begin{eqnarray}
 \zeta_\perp(0)=\frac{|\vec p_{K_2^*}|}{m_{K_2^*}}T_1^{BK_2^*}(0)= (0.29\pm0.09),\;\;\;
 \zeta_{||}(0)=\frac{1}{1-\frac{m_{K_2^*}^2}{m_BE_{K_2^*}}}\left(\frac{|\vec
 p_{K_2^*}|}{m_{K_2^*}}A_0^{BK_2^*}(0)-\frac{m_{K_2^*}}{m_B}\zeta_\perp(0)\right)=
 (0.26\pm0.10),\nonumber
\end{eqnarray}
show that they are of similar size, the same conclusion with the
$B\to V$ cases. This is not accidental but instead is an outcome of
the similar shapes between the vector and tensor meson LCDAs in the
dominant region of the PQCD approach. Our result is also accordance
with the theoretical estimate in Ref.~\cite{Hatanaka:2009gb}
\begin{eqnarray}
 \zeta_\perp(0)=0.27\pm0.03^{+0.00}_{-0.01}.
\end{eqnarray}

On the experimental side the branching ratio of the color-allowed
tree-dominated processes $B^0\to a_2^\pm \pi^\mp$ has been set with
an upper limit
\begin{eqnarray}
 {\cal B}(B^0\to a_2^\pm \pi^\mp)<3.0\times 10^{-4}.
\end{eqnarray}
When factorization is adopted this mode can be  used  to extract the
$B\to a_2$ form factor
\begin{eqnarray}
 |A_0^{B\to a_2}(q^2=0)|&<&7.6 F_+^{B\to \pi}(q^2=0)\simeq1.9,
\end{eqnarray}
where penguin contributions have been neglected as a result of their
small Wilson coefficients. Unfortunately the above constraint is too
loose to provide any useful information on the characters of the
tensor mesons. We expect more news on this front from the $B$
factories and other experiment facilities, including the Large
Hadron Collider.

At the leading order of $\alpha_s$, both $B\to K^*\gamma$ and $B\to
K_2^*\gamma$ only receive contributions from the chromo-magnetic
operator $O_{7\gamma}$, which leads to
\begin{eqnarray}
 {\cal B}(B\to K^*\gamma)&=& \tau_{B}\frac{G_F^2 \alpha_{\rm
 em}m_B^3m_b^2}{32\pi^4}\left(1-\frac{m_{K^*}^2}{m_B^2}\right)^3|V_{tb}V_{ts}^*C_7T_1^{BK^*}(0)|^2,\nonumber\\
 {\cal B}(B\to K_2^*\gamma)&=& \tau_{B}\frac{G_F^2 \alpha_{\rm
 em}m_B^5m_b^2}{256\pi^4
 m_{K_2^*}^2}\left(1-\frac{m_{K^*_2}^2}{m_B^2}\right)^5|V_{tb}V_{ts}^*C_7T_1^{BK^*_2}(0)|^2,
\end{eqnarray}
with $C_{7}$ being the Wilson coefficient for $O_{7\gamma}$ and
$V_{tb},V_{ts}$ being the CKM matrix element. Assuming that $C_7$ is
the same for the above two channels, we obtain the form factors
relation
\begin{eqnarray}
 \frac{ T_1^{BK_2^*}(0)}{ T_1^{BK^*}(0)}&=& (0.52\pm0.08),
\end{eqnarray}
from the experimental data~\cite{HFAG}
\begin{eqnarray}
 {\cal B}(B^-\to K^{*-}\gamma)&=& (42.1\pm1.8)\times 10^{-6},\;\;\;
 {\cal B}(B^-\to K^{*-}_2\gamma)=(14.5\pm4.3)\times 10^{-6}.
\end{eqnarray}
Our result for this ratio, roughly 0.7, is larger than this value
but is consistent with it when hadronic uncertainties from the final
mesons are taken into account. It also confirms our results that the
$B\to K_2^*$ form factors are smaller than the $B\to K^*$ ones at
$q^2=0$ point, in contrast to the LFQM results
$T_1^{BK_2^*}(0)\simeq T_1^{BK^*}(0)$~\cite{Cheng:2009ms}.

\section{Semilteptonic $B\to Tl\bar\nu$ decays}
\label{sec:semileptonic}

Integrating out the off shell W boson, one obtains the effective
Hamiltonian responsible for $b\to ul\bar \nu_l$ transition
 \begin{eqnarray}
 {\cal H}_{\rm eff}(b\to ul\bar \nu_l)=\frac{G_F}{\sqrt{2}}V_{ub}\bar
 u\gamma_{\mu}(1-\gamma_5)b \bar l\gamma^{\mu}(1-\gamma_5)\nu_l,
 \end{eqnarray}
where $V_{ub}$ is the CKM matrix element. In semileptonic $B\to
Tl\bar\nu_l$ decays, the helicity of the tensor meson can be
$h=0,\pm1$ but the $h=2$ configuration is not allowed physically.
Using the form factors obtained in the previous section, we can
investigate the semileptonic $B\to Tl\bar\nu$  decays with the
partial decay width
\begin{eqnarray}
 \frac{d\Gamma}{dq^2}&=&\sum_{i=L,\pm}
 \frac{d\Gamma_{i}}{dq^2},\nonumber\\
 \frac{d\Gamma_{L, \pm}}{dq^2}&=&
 \frac{ |G_F V_{ub}|^2  \sqrt  {\lambda_T} }{256m_{B_s}^3\pi^3q^2}
\left(1- {\frac{m_\ell^2}{q^2}}\right)^2
 (X_{L}, X_{\pm})
\end{eqnarray}
where $\lambda_T=\lambda(m^2_{B},m^2_T, q^2)$, and
$\lambda(a^2,b^2,c^2)=(a^2-b^2-c^2)^2-4b^2c^2$. The subscript
$(L,\pm)$ denotes the three polarizations of the tensor meson along
its momentum direction: $(0,\pm1)$.  $m_l$ represents the mass of
the charged lepton, and $q^2$ is the momentum square of the lepton
pair. In terms of the angular distributions, we can study the
forward-backward asymmetries (FBAs) of lepton which are defined as
 \begin{eqnarray}
 \frac{dA_{FB}}{dq^2} &=& \frac{\int^{1}_{0} dz (d\Gamma/dq^2dz) -
\int^{0}_{-1} dz (d\Gamma/dq^2dz)}{\int^{1}_{0} dz (d\Gamma/dq^2dz)
+ \int^{0}_{-1} dz (d\Gamma/dq^2dz)}\nonumber
 \end{eqnarray}
where $z=\cos\theta$ and the angle $\theta$ is the polar angle of
lepton with respect to the moving direction of the tensor meson in
the lepton pair rest frame. Explicitly, we have
\begin{eqnarray}
 \frac{dA_{\rm FB}}{dq^2}&=&
 \frac{1}{X_L + X_{+} + X_{-}} \left(\frac{\lambda_T}{6m_T^2m_B^2}  2  m_\ell^2
 \sqrt{\lambda_{T}}h_0(q^2) A_0(q^2)  -\frac{\lambda_T}{8m_T^2m_B^2}  4q^4\sqrt {\lambda_{T}} A_1(q^2) V(q^2)
 \right),\label{eq:AAS-tensor}
\end{eqnarray}
where
\begin{eqnarray}
X_{L} &=& \frac{2}{3}\frac{\lambda_T}{6m_T^2m_B^2} \left[
(2q^2+m_\ell^2)  h_0^2(q^2)  + 3
{\lambda_{T}} m_\ell^2 A_0^2 (q^2)\right]\,, \nonumber \\
X_{\pm} &=& \frac{2q^2}{3}  (2 q^2 + m_\ell^2
)\frac{\lambda_T}{8m_T^2m_B^2} \left[(m_{B}+m_{T})A_1(q^2) \mp
\frac{\sqrt{\lambda_{T}}
}{m_{B}+m_{T}}V(q^2) \right]^2,\nonumber\\
 h_0(q^2)&=& \frac{ 1}{2 m_{T}
}\left[(m_{B}^2-m_{T}^2-q^2)(m_{B}+m_{T})A_1(q^2)-
\frac{{\lambda_{T}
}}{m_{B}+m_{T}}A_2(q^2)\right].\label{eq:longitudinal-Ds1}
\end{eqnarray}
Integrating over the $q^2$, we obtain the partial decay width and
integrated angular asymmetry  for this decay mode
\begin{eqnarray}
 \Gamma=\Gamma_L+\Gamma_++\Gamma_-,\;\;\;
 A_{FB}  &=& \frac{1}{\Gamma}\int dq^2 \int^{1}_{-1}sign(z) dz (d\Gamma/dq^2dz) \nonumber
\end{eqnarray}
with $
 \Gamma_{L,\pm}=\int_{m_l^2}^{(m_{B}-m_{T})^2} dq^2
 \frac{d\Gamma_{L,\pm}}{dq^2}.$
Physical quantities ${\cal B}_{\rm{L}}$, ${\cal B}_{\rm{+}}$, ${\cal
B}_{\rm{-}}$, and ${\cal B}_{\rm{total}}$  can be obtained through
different experimental measurements, where ${\cal B}_{\rm{T}}={\cal
B}_{\rm{+}}+{\cal B}_{\rm{-}}$ and ${\cal B}_{\rm{total}}={\cal
B}_{\rm{L}}+{\cal B}_{\rm{T}}$ with ${\cal B}_{\rm{L}}$, ${\cal
B}_{\rm{+}}$ and ${\cal B}_{\rm{-}}$ corresponding to contributions
of different polarization configurations to branching ratios. Since
there are three different polarizations,  it is also meaningful to
define the polarization fraction
\begin{eqnarray}
 f_L=\frac{\Gamma_L}{\Gamma_L+\Gamma_++\Gamma_-}.
\end{eqnarray}

Our theoretical results for the $B\to T l\bar\nu_l$ ($l=e,\mu$) and
$B\to T \tau\bar\nu_{\tau}$ decays are listed in Table
\ref{tab:branchratios1}, with masses of the electron and muon
neglected in the case of $l=e,\mu$. The $B$ meson lifetime is taken
from the particle data group and the CKM matrix element $V_{ub}$ is
employed as $|V_{ub}|=(3.89\pm0.44)\times
10^{-3}$~\cite{Amsler:2008zz}.

Some remarks are given in order.
 \begin{itemize}
\item Most of the total branching ratios are of the order $10^{-4}$,
implying a promising prospect to measure these channels at the Super
B factories and the LHCb. A tensor meson can be reconstructed in the
final state of  two or three pseudoscalar mesons.

\item  The
heavy $\tau$ lepton will bring a smaller phase space than the
lighter electron, thus the branching ratios of $B\to T
\tau\bar\nu_{\tau}$ decays are smaller than those of the
corresponding $B\to T e\bar\nu_e$ decay modes by a factor of 3.

\item
The positively polarized branching ratio  ${\rm Br}_+$ is tiny since
the $A_1$ term cancels with the contribution from $V$. The
longitudinal contributions are about twice as large as the
transverse polarizations and accordingly the polarization fraction
$f_L$ is around $(60\%-70\%)$

\item For $l=(e,\mu)$ the angular asymmetries are negative since only the
second terms in Eq.~\eqref{eq:AAS-tensor} contribute. In the case of
$l=\tau$, the two terms give destructive contributions, resulting in
tiny angular asymmetries in magnitude.

\item In the polarization fractions and angular asymmetries, the
uncertainties from the form factors and the CKM matrix element will
mostly cancel and thus they are stable against hadronic
uncertainties.
 \end{itemize}

\begin{table}
\caption{The branching ratios, polarizations and angular asymmetries
for the $b\to ul\bar \nu_l$ ($l=e,\mu$) and $b\to u\tau\bar
\nu_\tau$ decay channels (in units of $10^{-4}$). ${\cal
B}_{\rm{L}}$ and ${\cal B}_{\pm}$ are the longitudinally and
transversely polarized contributions to the branching ratios.    }
 \label{tab:branchratios1}
 \begin{center}
 \begin{tabular}{ccccccc}
 \hline\hline
 \ \ \      & ${\cal
B}_{\rm{L}}$  &${\cal B}_+$  &${\cal B}_-$   &${\cal
B}_{\rm{total}}$  &$f_L$ & $A_{\rm FB}$\\
 \hline
 \ \ \ $\overline B^0\to a_2^+l\bar
 \nu_l$         &$   0.85
            _{  -0.42         }
            ^{+   0.60        }
 $    &$\sim0.007$  & $   0.30
            _{  -0.15         }
            ^{+   0.21        }
 $&$1.16 ^{+0.81}_{-0.57} $  & $
 73.3
             _{  -0.5         }
            ^{+   0.4        }
 $ & $-0.186^{+0.003}_{-0.002}$  \\
 \hline
 \ \ \ $B^-\to f_2^0 (1270)l\bar
 \nu_l$   & $   0.52
             _{  -0.26         }
            ^{+   0.36        }
 $    & $\sim 0.004$  & $   0.17
             _{  -0.08         }
            ^{+   0.11        }
 $  &$0.69 _{-0.34}^{+0.48} $  &$
 74.9
             _{  -0.7         }
            ^{+   0.6        }
 $ & $-0.175\pm0.004$  \\
 \hline
 \ \ \ $\overline B_s\to K_2^{*+}(1430)l\bar
 \nu_l$    &$   0.50
             _{  -0.23         }
            ^{+   0.32        }
 $    &$\sim0.006$  &$   0.23
             _{  -0.11         }
            ^{+   0.15        }
 $  &$0.73 ^{+0.48}_{-0.33} $  &$
 68.3
             _{  -0.5         }
            ^{+   0.5        }
 $ &$-0.221\pm0.005$   \\
 \hline\hline
 \ \ \ $\overline B^0\to a_2^+\tau\bar \nu_{\tau}$         &$   0.29
             _{  -0.14         }
            ^{+   0.20        }
 $   &$\sim 0.004$  &$   0.12
             _{  -0.06         }
            ^{+   0.08        }
 $ &$0.41 ^{+0.29}_{-0.20} $  & $  69.9
             _{  -0.6         }
            ^{+   0.6        }
 $ &$0.031\pm0.005$ \\
 \hline
 \ \ \ $B^-\to f_2^0(1270)\tau\bar \nu_{\tau}$   & $   0.18
             _{  -0.09         }
            ^{+   0.13        }
 $    &$\sim 0.002$  & $   0.07
             _{  -0.03         }   ^{+   0.05        }
 $
       &$0.25 ^{+0.18}_{-0.13} $  &  $
       71.9
             _{  -0.8         }
            ^{+   0.8        }
 $ & $0.048\pm0.007$   \\
 \hline
 \ \ \ $\overline B_s\to K_2^{*+}(1430)\tau\bar \nu_{\tau}$    &$   0.16
             _{  -0.07         }
            ^{+   0.10        }
 $  &$\sim 0.003$  &$   0.09
             _{  -0.04         }
            ^{+   0.06        }
 $  &$0.25 ^{+0.17}_{-0.12})$  & $  64.1
             _{  -0.3         }
            ^{+   0.4        }
 $ & $-0.024\pm0.005$  \\
 \hline\hline
 \end{tabular}
 \end{center}
 \end{table}
\section{Summary}

Inspired by the success of the PQCD approach in the application to
$B$ decays into s-wave mesons, we give a comprehensive study on the
$B\to T$ transition form factors. Our results will become necessary
inputs in the analysis of the nonleptonic $B$ decays into a tensor
meson.

The similarities in the Lorentz structures of the wave functions and
$B$ decay form factors involving a vector and a tensor meson allow
us to obtain the factorization formulas of  $B\to T$ form factors
from the $B\to V$ ones. Furthermore, the light-cone distribution
amplitudes of tensor mesons and vector mesons have similar shapes in
the dominant region of the perturbative QCD approach, and thus these
two sets of form factors are found to have the same signs and
related $q^2$-dependence behaviors. In the large recoil region, we
find that our results for the form factors satisfy the relations
derived from the large energy limit. The two independent functions
$\zeta_\perp$ and $\zeta_{||}$ are found to have similar size at
$q^2=0$ point. We also find that the $B\to T$ form factors are
smaller than the $B\to V$ ones, which is supported by the
experimental data of radiative $B$ decays.

At last, we also use these results to explore semilteptonic $B\to
Tl\bar \nu_l$ decays and we find that the branching fractions can
reach the order $10^{-4}$, implying a promising prospect to observe
these channels.

\section*{Acknowledgements}

I would like to acknowledge  Hai-Yang Cheng and Ying Li for useful
discussions. I am very grateful to Pietro Colangelo and Fulvia De 
Fazio for their warm hospitality during my stay in Bari. This work is supported by the INFN through the program of INFN fellowship for foreigners and also 
in part by the National Natural Science Foundation of
China under the Grant No. 10805037 and 10947007.


\end{document}